\documentclass[twocolumn]{autart}

\usepackage[square, comma, sort]{natbib}            
\setcitestyle{numbers}

\usepackage{amsmath,amssymb,amsfonts}
\usepackage{algorithm}
\usepackage{algorithmicx}
\usepackage{algpseudocode}
\usepackage{graphicx} 
\usepackage{textcomp}
\usepackage{xcolor}
\usepackage{caption}
\usepackage{array}
\usepackage{float}
\usepackage{dsfont}
\usepackage{hyperref}

\usepackage[textwidth=1.5in]{todonotes}

\renewcommand{\thetable}{\arabic{table}}
\def\BibTeX{{\rm B\kern-.05em{\sc i\kern-.025em b}\kern-.08em
    T\kern-.1667em\lower.7ex\hbox{E}\kern-.125emX}}    
\usepackage{etoolbox}

\newtheorem{definition}{Definition}
\newtheorem{remark}{Remark}
\newtheorem{theorem}{Theorem}

\DeclareMathOperator{\vect}{vec}
\algnewcommand{\LeftComment}[1]{\Statex \(\triangleright\) #1}
\newfloat{algorithm}{t}{lop}

\begin{document}
\begin{frontmatter}
\title{A Barrier Function Approach to Finite-Time Stochastic System Verification and Control}
\author[santoyodutreix]{Cesar Santoyo}, \author[santoyodutreix]{Maxence Dutreix}, and \author[coogan]{Samuel Coogan
}
\thanks[santoyodutreix]{C. Santoyo ({\tt csantoyo@gatech.edu}) and M. Dutreix ({\tt maxdutreix@gatech.edu}) are with the School of Electrical \& Computer Engineering, Georgia Institute of Technology, Atlanta, GA, 30318, USA.}
\thanks[coogan]{S. Coogan ({\tt sam.coogan@gatech.edu}) is with the School of Electrical \& Computer Engineering and the School of Civil and Environmental Engineering, Georgia Institute of Technology, Atlanta, GA, 30318, USA.}
\thanks{This work was partially supported by NSF under Grant \#1749357. C. Santoyo was supported by the NSF Graduate Research Fellowship Program under Grant No. DGE-1650044.}

\begin{abstract}
  This paper studies the problem of enforcing safety of a stochastic dynamical system over a finite-time horizon. We use stochastic control barrier functions as a means to quantify the probability that a system exits a given safe region of the state space in finite time. A barrier certificate condition that bounds the expected value of the barrier function over the time horizon is recast as a sum-of-squares optimization problem for efficient numerical computation. Unlike prior works, the proposed certificate condition includes a state-dependent upper bound on the evolution of the expectation. We present formulations for both continuous-time and discrete-time systems. Moreover, for systems for which the drift dynamics are affine-in-control, we propose a method for synthesizing polynomial state feedback controllers that achieve a specified probability of safety. Several case studies are presented which benchmark and illustrate the performance of our verification and control method in the continuous-time and discrete-time domains.
\end{abstract}
\end{frontmatter}

\section{Introduction}
Reliance on complex, safety-critical systems is increasing, which has made safety verification of such systems of utmost importance. For example, environments populated by both humans and autonomous systems (e.g. fulfillment centers and autonomous vehicles) require rigorous safety verification to ensure desired behavior is achieved. From a practical standpoint, safety verification can translate directly to ensuring qualitative guidelines such as collision avoidance are maintained. Safety-critical systems are often analyzed in a purely deterministic framework, however, many real-world applications are subject to stochastic disturbances and are better modeled as stochastic systems.  \par 
A common approach to safety verification in deterministic systems is via \textit{barrier functions} which provide Lyapunov-like guarantees regarding system behavior. The existence of a barrier function which satisfies a \textit{barrier certificate} can often be enough to certify the safe operation of a system \cite{PrajnaStoch2007}.
Recent work has modified and improved the deterministic form of barrier functions and expanded their application. In particular, control barrier functions have been introduced to guarantee safety of affine-in-control systems \cite{wieland2007constructive, AmesAaronD.2017CBFB}.  This is demonstrated in applications for cruise control \cite{AmesAaronD.2017CBFB, ames2014control}, collision avoidance in robotic swarms \cite{WangLi2017SBCf}, walking robots \cite{Hsu:2015uo}, and has recently been extended to allow for input-to-state safe control barrier functions \cite{KolathayaShishir2018ISWC} and to guarantee
finite-time convergence to a safe region
\cite{AnqiLi}. \par 
In the stochastic setting, continuous-time (CT) safety verification via barrier certificates for infinite time horizons was introduced in \cite{PrajnaStoch2007} alongside the deterministic counterpart. The work presented in \cite{PrajnaStoch2007} provides a framework for bounding the probability a system will ever exit a safe region based on a non-negative barrier function defined on the system state space. 
To obtain probabilistic guarantees over infinite time horizons, \cite{Prajna:2004up} requires the infinitesimal generator, which dictates the expected value evolution of a stochastic process, to be non-positive; i.e., the barrier function is required to be a \textit{supermartingale}.\par

The paper \cite{SteinhardtJacob2012Frvo} relaxes the supermartingale condition for finite-time safety verification and instead provides a barrier certificate which only requires the infinitesimal generator of the barrier process to be upper bounded by a constant. Such processes are called \textit{c-martingales} and allow the expected value of the barrier function to increase over time. This approach results in a safety probability bound for finite-time horizons.\par
Recently, discrete-time (DT) control barrier functions have been used to certify safety for bi-pedal robots \cite{agrawal2017discrete}, safe policy synthesis for multi-agent systems \cite{ahmadi2019safediscretetime} and for temporal logic verification of discrete-time systems \cite{Pushpak_Sadegh_Zamani_2018, jagtap2019formal}. \par 
The work presented in \cite{agrawal2017discrete} mirrors, in discrete-time, the formulation of deterministic continuous-time control barrier functions initially presented in \cite{ames2014control} whose overarching theory and applications are summarized in \cite{AmesCooganEgerstedtNotomistaSreenathTabuada2019contro}. In \cite{agrawal2017discrete}, the formulation for discrete-time barrier functions presents a significant distinction from the continuous-time counterpart resulting in a nonlinear optimization problem which is not necessarily convex. This poses challenges in solving the stochastic discrete-time controller synthesis problem in a similar manner to that of stochastic continuous systems shown in \cite{Santoyo2019VerificationAC}. There exist few publications related to verification and control of stochastic discrete-time systems. \par 
\par 
The present paper studies the problem of verifying safety of stochastic systems on finite time horizons for both continuous-time and discrete-time domains, and the contributions are as follows. We build on the approaches proposed in \cite{PrajnaStoch2007, SteinhardtJacob2012Frvo} and propose a barrier certificate constraint that imposes a state-dependent bound on the expected value for both continuous-time and discrete-time systems. This bound was originally proposed and studied by Kushner in \cite{Kushnerstochstabtext, KushnerH.1966Ftss, kushner1971introductiontostochcontrol} in the context of stochastic stability. The proposed barrier certificate allows the expected value of the barrier to increase and covers the c-martingale condition of \cite{SteinhardtJacob2012Frvo} as a special case.  However, our formulation also accounts for the system dynamics in the expectation constraint. This allows for probability bounds that are no worse than the c-martingale condition, and in many cases, especially with high values of sigma, provides better probability bounds. 
\par 
As in \cite{PrajnaStoch2007, SteinhardtJacob2012Frvo}, we compute barrier functions using \textit{sum-of-squares} (SOS) optimization. Like in \cite{PrajnaStoch2007}, but unlike \cite{SteinhardtJacob2012Frvo}, we utilize polynomial barrier functions. This provides a simpler formulation of the probability of failure on a finite time horizon when compared to the approach in \cite{SteinhardtJacob2012Frvo} which uses exponential barrier functions and, empirically, provides tighter probability bounds. \par 
Third, we extend our formulation to allow for control inputs and provide a method for synthesizing a safe controller. In particular, we consider affine-in-control systems and the proposed approach searches for a polynomial state feedback controller which ensures a system's failure probability achieves a predetermined criterion via a \emph{stochastic control barrier function}. \par 
Our preliminary work on continuous-time verification and control synthesis is published in \cite{Santoyo2019VerificationAC} with two case studies. The paper \cite{Santoyo2019VerificationAC} only focused on continuous-time. In this paper, we consider stochastic system verification and control law synthesis in the discrete-time setting.
\par 
This paper is organized as follows: Section \ref{section: prelims} covers the background information of stochastic differential and difference equations, barrier functions and SOS optimization. Section \ref{section: probform} presents the problem formulation. Section \ref{section: methodology} highlights the methodology we utilize to solve the SOS optimization and stochastic control problem.  Section \ref{section: casestudies} and Section \ref{section: conclusion} present numerical case studies which illustrate our results and conclusions, respectively. \par 
\section{Preliminaries}
\label{section: prelims}
In this section, we first introduce background information regarding stochastic systems, stochastic processes, and SOS polynomials.
\subsection{Stochastic Differential Equations}
Consider a complete probability space ($\Omega, \mathcal{F}, P$) and a standard Wiener process $w(t)$ taking values in $\mathbb{R}^m$. We consider continuous-time stochastic processes $x(t)$ satisfying a stochastic differential equation of the form
\begin{equation} \label{stochasticdiffeq}
	dx = F(x)dt + \sigma(x)dw
\end{equation}
 where the compact set $\mathcal{X} \subset \mathbb{R}^n$ is the system state space, $F: \mathcal{X} \rightarrow \mathbb{R}^n$ is the drift rate and $\sigma: \mathcal{X} \rightarrow \mathbb{R}^{n \times m}$ is the diffusion term. We assume the functions $F(x)$ and $\sigma(x)$ are Lipschitz continuous. 
 We now introduce the infinitesimal generator, which extends the usual definition of a time derivative to instead consider the expectation of a function of a random process \cite{ksendalB.K.BerntKarsten1998Sde:}.
\begin{definition}
Let $x(t)$ be a stochastic process in $\mathbb{R}^n$. The \emph{infinitesimal generator} $\mathcal{A}$ of $x(t)$ acts on functions of the state space and is defined as
\begin{equation}
\nonumber
\mathcal{A}B(x) = \lim\limits_{t \downarrow 0} \frac{\mathbb{E}[B(x) | x_0] - B(x_0)}{t}
\end{equation}
where $B: \mathcal{X} \rightarrow \mathbb{R}$ such that the limit exists for all $x_0=x(0)$. 
\end{definition} \par 
In particular, the infinitesimal generator for any process as in (\ref{stochasticdiffeq}) is of the form shown in Fact \ref{generatorfact}. 
\begin{fact}[{Ch. 7, Theorem 7.3.3 of \cite{ksendalB.K.BerntKarsten1998Sde:}}]
    \label{fact:infinitesimalgen}
	Let $x(t)$ be a stochastic process satisfying (\ref{stochasticdiffeq}), then the \emph{infinitesimal generator} $\mathcal{A}$ of some twice differentiable function $B(x)$ is given by
	\begin{equation}
		\nonumber
		\mathcal{A}B(x) =  \sum_{i = 1}^{n}F_i(x)\frac{\partial B }{\partial x_i} + \frac{1}{2}\sum_{i = 1}^{n}\sum_{j = 1}^{n}\bigg(\sigma(x) \sigma^T(x)\bigg)_{i,j}\frac{\partial^2B}{\partial x_i \partial x_j}.
	\end{equation}
	\vspace{-7mm}
	\label{generatorfact}
\end{fact}
\par The stochastic process $x(t)$ is not guaranteed to lie in $\mathcal{X}$ at all times which leads us to define the stopped process $\tilde{x}$.
\begin{definition}[\cite{PrajnaStoch2007}, Definition 12]
	Suppose that $\tau$ is the first time of exit of $x(t)$ from the open set Int($\mathcal{X}$). Then the \emph{stopped process} $\tilde{x}(t)$ is defined by
	\begin{equation} \nonumber
	\tilde{x}(t) = 
		\begin{cases} 
		x(t) & \text{for} \ t < \tau \\
		x(\tau) & \text{for} \ t \geq \tau.
		\end{cases}
	\end{equation}
	\vspace{-7mm}
	\label{definition: CTstoppedproc}
\end{definition}
The stopped process $\tilde{x}(t)$ inherits the same strong Markovian property of $x(t)$ and shares the same infinitesimal generator \cite{Kushnerstochstabtext}.
\subsection{Stochastic Difference Equations}
Consider now a discrete-time stochastic process of the form \cite{kushner1971introductiontostochcontrol}
\begin{equation}
    \label{stochdifferenceeq}
    x[k+1] = F(x[k]) + \sigma(x[k])\xi[k]
\end{equation}
where $\mathcal{X} \subset \mathbb{R}^n$, 
$\xi[k] \in  \mathbb{R}^p$, $F:\mathbb{R}^n \rightarrow \mathbb{R}^n$, and $\sigma:\mathbb{R}^{n} \rightarrow \mathbb{R}^{n \times p}$. Here, $\xi[k]$ is a random disturbance whose value is governed by some distribution at each time step $k$. 
For the discrete-time setting, a stopped process is defined analogously to Definition \ref{definition: CTstoppedproc} and denoted by $\tilde{x}[k]$.
\subsection{Sum-of-Squares}
\begin{definition}
	\label{sumofsquaresdefinition}
	Define $\mathbb{R}[x]$ as the set of all polynomials in $x\in\mathbb{R}^n$. Then
	\begin{equation}
		\nonumber
		\Sigma[x] \triangleq \bigg\{   s(x) \in \mathbb{R}[x] : s(x) = \sum_{i=1}^{m} g_i(x)^2, g_i(x) \in \mathbb{R}[x] \bigg\}
	\end{equation}
	is the set of \emph{sum-of-squares polynomials.} 
\end{definition}
Note that if $s(x) \in \Sigma[x]$ then $s(x) \geq 0$ $\forall$ $x$.
\begin{definition}
	Given $p_i(x) \in \mathbb{R}[x]$ for $i = 0, \ldots, m$, the problem of finding $q_i(x) \in \Sigma[x]$ for $i = 1, \ldots, \hat{m}$ and $q_i(x)\in\mathbb{R}[x]$ for $i=\hat{m}+1,\ldots, m$ such that
	\begin{equation}
		\nonumber
		p_0(x) + \sum_{i = 1}^{m}p_i(x)q_i(x) \in \Sigma[x]
	\end{equation}
	is a \emph{sum-of-squares program (SOSP).} 
\end{definition}
SOSPs can be efficiently converted to semidefinite programs using tools such as SOSTOOLS \cite{sostools}.
\section{Problem Formulation}
\label{section: probform}
We address the problem of creating a bound on the probability a stochastic system of form (\ref{stochasticdiffeq}) or (\ref{stochdifferenceeq}) exits a safe region during a finite-time horizon. Additionally, we present an algorithmic approach for control synthesis based on a system's probability of becoming unsafe. With this, we achieve the following objectives for both continuous-time and discrete-time systems.\par 
\textbf{Objectives (CT \& DT):}
(Verification) First, given a continuous-time or discrete-time stochastic system of the form (\ref{stochasticdiffeq}) or (\ref{stochdifferenceeq}) and a fixed time horizon, upper bound the probability of failure, \emph{i.e.}, the probability that the system's state reaches a set of unsafe conditions within the finite time horizon. (Synthesis) Second, given a continuous-time or discrete-time stochastic system with input, synthesize a feedback control law to achieve a desired maximum probability of failure. \par 
\subsection{Continuous Time Systems}
Consider the stochastic process $x(t)$ which satisfies the stochastic differential equation
\begin{equation}
\label{controlstochdiff}
dx = (f(x) + g(x)u(x) )dt + \sigma(x)dw
\end{equation}
where $f:\mathcal{X} \rightarrow \mathbb{R}^n $, $g:\mathcal{X} \rightarrow \mathbb{R}^{n \times p} $, $\sigma:\mathcal{X} \rightarrow \mathbb{R}^{n \times m}$ and  $w$ is a $m$-dimensional Wiener process. Additionally, $u: \mathcal{X} \rightarrow \mathbb{R}^p$ where $u$ is a state feedback control law. We define $F(x) = f(x) + g(x)u(x)$. In the derivation below, we consider $u(x)$ given and fixed and hence $F$ is a function only of $x$. In Section \ref{subsection: VerControlLawSynth}, when we address the problem of synthesizing a feedback control law $u(x)$, it is then implicit that $F$ depends on this choice of feedback.
The following theorem is an immediate corollary of \cite[Chapter 3, Theorem 1]{Kushnerstochstabtext} and recovers the supermartingale condition \cite[Theorem 15]{PrajnaStoch2007} and c-martingale condition \cite[Theorem 2.4]{SteinhardtJacob2012Frvo} as special cases.
\begin{theorem} \label{theorem:SBF_CT}
	Given the stochastic differential equation (\ref{controlstochdiff}) and the sets $\mathcal{X} \subset \mathbb{R}^n$, $\mathcal{X}_u \subseteq \mathcal{X}, \mathcal{X}_0 \subseteq \mathcal{X}\setminus \mathcal{X}_u$ with $F(x) = f(x) + g(x)u(x)$ and $\sigma(x)$ locally Lipschitz continuous, where $u(x)$ is some feedback control law. Consider the stopped process $\tilde{x}(t)$. Suppose there exists a twice differentiable function $B$ such that
	\begin{align}
	    \label{eq: CTconstraint1}
    	B(x) &\leq \gamma \ \forall x \in \mathcal{X}_0 \\
	    \label{eq: CTconstraint2}
    	B(x) &\geq 1 \ \forall x \in \mathcal{X}_u \\
	    \label{eq: CTconstraint3}
    	B(x) &\geq 0 \ \forall x \in \mathcal{X}
    	\\
    	\nonumber
    	\frac{\partial B}{\partial x}F(x) + \frac{1}{2}&\text{Trace}\bigg(\sigma ^T(x) \frac{\partial^2B}{\partial x^2}\sigma (x) \bigg)  \\
    	\label{supmartrelax}  \leq -\alpha B(x) &+ \beta \ \forall x \in \mathcal{X}\setminus \mathcal{X}_u
	\end{align}
	for some $\alpha \geq 0$, $\beta \geq 0$ and $\gamma \in [0, 1)$. Define
        \begin{align}
          \label{eq:2}
          \rho_{u}&:=P \{\tilde{x}(t)\in \mathcal{X}_u \ \text{ for } \ 0\leq t\leq T \ | \ \tilde{x}(0) \in \mathcal{X}_0\} \\
           \rho_{B}&:= P\left\{\sup_{0\leq t \leq T} B\big(\tilde{x} \big) \geq 1 \ | \ \tilde{x}(0) \in \mathcal{X}_0 \right\}.
        \end{align}
Then 
\begin{itemize}
\item If $\alpha > 0$ and $\frac{\beta}{\alpha} \leq 1$,
    \vspace{-3.5mm}
	\begin{equation}
	\label{bound1}
\rho_u \leq \rho_{B} \leq 1 - \bigg(1 - \gamma \bigg)e^{-\beta T}.
	\end{equation}
\item If $\alpha > 0$ and $\frac{\beta}{\alpha} \geq 1$,
    \vspace{-3.5mm}
	\begin{equation}
	\label{bound2}
\rho_u \leq \rho_{B} \leq \frac{\gamma + (e^{\beta T} - 1)\frac{\beta}{\alpha}}{e^{\beta T}}.
	\end{equation}
\item If $\alpha = 0$,
    \begin{equation}
	\label{bound3}
\rho_u \leq \rho_{B} \leq \gamma + \beta T.
      \end{equation}
    \end{itemize}
  \end{theorem} \par
The bound (\ref{bound3}) is characterized in \cite{Pushpak_Sadegh_Zamani_2018} and \cite{SteinhardtJacob2012Frvo} as the upper bound on the probability of being unsafe for a c-martingale. \par
If $B(x)$ satisfies the conditions of Theorem \ref{theorem:SBF_CT}, then $B(x)$ is called a \emph{stochastic control barrier function} for a given control policy $u(x)$. Relaxing the supermartingale condition on the infinitesimal generator in the fashion of Theorem \ref{theorem:SBF_CT} gives three case-dependent finite time probability bounds on a system's likelihood of entering an unsafe region in the form of (\ref{bound1}), (\ref{bound2}), and (\ref{bound3}).
\begin{remark}
  \label{rem:1}
If the initial state $x_0$ is known exactly,
 then $B(x_0)$ can be substituted for $\gamma$ in the probability bounds of Theorem \ref{theorem:SBF_CT}. This provides an upper bound on the probability of failure over a particular initial point rather than on an initial set, $\mathcal{X}_0$.   
\end{remark}
\subsection{Discrete Time Systems}
Consider the stochastic discrete-time system 
\begin{equation} \label{discretetime}
    x[k+1] = f(x[k]) + g(x[k])u(x[k]) + \sigma(x[k])\xi[k]
\end{equation}
where $f:\mathcal{X} \rightarrow \mathbb{R}^n $, $g:\mathcal{X} \rightarrow \mathbb{R}^{n \times p} $, $\sigma:\mathcal{X} \rightarrow \mathbb{R}^{n \times m}$ and  $\xi$ is a stochastic process whose value is governed by some probabilistic distribution. Additionally, $u: \mathcal{X} \rightarrow \mathbb{R}^p$ where $u(x)$ is a polynomial control law. We define $F(x, \xi) = f(x) + g(x)u(x) + \sigma(x)\xi$. The following theorem is an immediate corollary of  \cite[Chapter 3, Theorem 3]{Kushnerstochstabtext}.
\begin{theorem} \label{theorem:SBF_DT}
	Given the stochastic difference equation (\ref{discretetime}) and the sets $\mathcal{X} \subset \mathbb{R}^n$, $\mathcal{X}_u \subseteq \mathcal{X}, \mathcal{X}_0 \subseteq \mathcal{X}\setminus \mathcal{X}_u$ with $F(x, \xi) = f(x) + g(x)u(x) + \sigma(x)\xi$ where $u(x)$ is some feedback control law. Consider the stopped process $\tilde{x}[k]$. Suppose there exists a twice differentiable function $B$ such that
	\begin{align}
	B(x) &\leq \gamma \ \forall x \in \mathcal{X}_0 \\
	B(x) &\geq 1 \ \forall x \in \mathcal{X}_u \\
	B(x) &\geq 0 \ \forall x \in \mathcal{X}
	\end{align}
	\begin{equation} \label{expectationevolbound}
	\mathbb{E}[B(F(x,\xi))\ |\ x] \leq \frac{B(x)}{\tilde{\alpha}} + \tilde{\beta}  \ \ \ \forall x \in \mathcal{X}\setminus \mathcal{X}_u
	\end{equation}
	for some $\tilde{\alpha} \geq 1$, $0 \leq \tilde{\beta}< 1$ and $\gamma \in [0, 1)$. Define
        \begin{align}
          \label{eq:2}
          \rho_{u}&:=P \{\tilde{x}[k] \in \mathcal{X}_u \text{ for } 0\leq k \leq N \ | \ \tilde{x}[0] \in \mathcal{X}_0\} \\
           \rho_{B}&:= P\left\{\sup_{0\leq k \leq N} B(\tilde{x}) \geq 1 \ | \ \tilde{x}[0] \in \mathcal{X}_0 \right\}.
        \end{align}
Then 
\begin{itemize}
\item If $\tilde{\alpha} > 1$ and $\frac{\tilde{\beta}\tilde{\alpha}}{\tilde{\alpha} - 1} \leq 1$,
\vspace{-3.5mm}
\begin{equation}
	\label{disbound1}
    \begin{split}
       \rho_u \leq \rho_{B}
        \leq 1 - \bigg(1-\gamma\bigg)\prod\limits^{N-1}_0\bigg(1 - \tilde{\beta} \bigg).
    \end{split}
\end{equation}
\item If $\tilde{\alpha} > 1$ and $\frac{\tilde{\beta}\tilde{\alpha}}{\tilde{\alpha} - 1} > 1$,
    \vspace{-3.5mm}
	\begin{equation} 
	\label{disbound2}
	\begin{split}
	    \rho_u \leq \rho_{B}
	    \leq \gamma\tilde{\alpha}^{-N} + \frac{(1 - \tilde{\alpha}^{-N})\tilde{\alpha}\tilde{\beta}}{(\tilde{\alpha} - 1)}.
	\end{split}
	\end{equation}
\item If $\tilde{\alpha} = 1$,
    \vspace{-3.5mm}
    \begin{equation} \label{disbound3}
        \begin{split}
            \rho_u \leq \rho_{B} \leq \gamma + \tilde{\beta}N. 
        \end{split}
    \end{equation}
    \end{itemize}
  \end{theorem} \par
  
Like in continuous time, if $B(x)$ satisfies the conditions of Theorem \ref{theorem:SBF_DT}, then $B(x)$ is called a stochastic control barrier function for a given control policy $u(x)$. Additionally, like in continuous-time, Remark \ref{rem:1} also applies.
\section{SOS Formulations \& Numerical Procedures}
\label{section: methodology}
In this section we present our approach to construct both continuous-time and discrete-time stochastic control barrier functions based on the problem formulations of Section \ref{section: prelims}. First, we adapt the inequality constraints given in Theorem \ref{theorem:SBF_CT} \& \ref{theorem:SBF_DT} to be formulated as an SOSP when $\alpha$ and $u(x)$ are known. Second, we present the algorithms which construct barrier functions and present our method for computing a control policy. \par
\subsection{SOS Formulation for Safety Verification}
For continuous-time system verification, the conditions in Theorem \ref{theorem:SBF_CT} can be recast as SOS constraints. 
\begin{theorem} \label{SOSPTheoremCT}
    Consider a system of the form of (\ref{controlstochdiff}) and the sets $\mathcal{X}$, $\mathcal{X}_0$, and $\mathcal{X}_u$ and assume these sets are described as $\mathcal{X} = \{x\in \mathbb{R}^n : s_{\mathcal{X}}(x) \geq 0 \}$, $\mathcal{X}_0 = \{x\in \mathbb{R}^n : s_{\mathcal{X}_o}(x) \geq 0 \}$, and $\mathcal{X}_u = \{x\in \mathbb{R}^n : s_{\mathcal{X}_u}(x) \geq 0 \}$ for some polynomials $s_{\mathcal{X}}$, $s_{\mathcal{X}_o}$, and $s_{\mathcal{X}_u}$. Suppose there exists a polynomial $B(x)$, and SOS polynomials $\lambda_{\mathcal{X}}(x)$, $\lambda_{\mathcal{X}_o}(x)$, and $\lambda_{\mathcal{X}_u}(x)$ that satisfy 
    \begin{align} 
        \label{CTSOSP_constraint1}
        B(x) - \lambda_{\mathcal{X}}(x)s_{\mathcal{X}}(x) &\in \Sigma[x] \\
        \label{CTSOSP_constraint2}
        B(x) - \lambda_{\mathcal{X}_u}(x)s_{\mathcal{X}_u}(x) - 1 &\in \Sigma[x] \\
        \label{CTSOSP_constraint3}
        -B(x) - \lambda_{\mathcal{X}_o}(x)s_{\mathcal{X}_o}(x) + \gamma &\in \Sigma[x] \\
        \nonumber
        -\frac{\partial B(x)}{\partial x}F(x)  -\frac{1}{2}\text{Trace}\bigg(\sigma ^T(x) \frac{\partial^2B}{\partial x^2}&\sigma (x) \bigg) - \alpha B(x) + \beta  \\ \label{CTSOSP_constraint4}-\lambda_{\mathcal{X}_u}(x)s_{\mathcal{X}_u}(x) - \lambda_{\mathcal{X}}(x)s_{\mathcal{X}}(x) &\in \Sigma[x] 
    \end{align}
    where $F(x) = f(x) + g(x)u(x)$. Then, the probability of failure, depending on the values of $\alpha$ and $\beta$, satisfies (\ref{bound1}), (\ref{bound2}) or (\ref{bound3}).
\end{theorem}
Theorem \ref{theorem:SBF_DT} for discrete-time verification for systems of the form of (\ref{discretetime}) can also be recast as SOS constraints.
\begin{theorem}
\label{SOSPTheoremDT}
    Consider a system of the form of (\ref{discretetime}) and the sets $\mathcal{X}$, $\mathcal{X}_0$, and $\mathcal{X}_u$ and assume these sets can be described as $\mathcal{X} = \{x\in \mathbb{R}^n : s_{\mathcal{X}}(x) \geq 0 \}$, $\mathcal{X}_0 = \{x\in \mathbb{R}^n : s_{\mathcal{X}_o}(x) \geq 0 \}$, and $\mathcal{X}_u = \{x \in \mathbb{R}^n : s_{\mathcal{X}_u}(x) \geq 0 \}$ for some polynomials $s_{\mathcal{X}}$, $s_{\mathcal{X}_o}$, and $s_{\mathcal{X}_u}$. Suppose there exists a polynomial $B(x)$, and SOS polynomials $\lambda_{\mathcal{X}}(x)$, $\lambda_{\mathcal{X}_o}(x)$, and $\lambda_{\mathcal{X}_u}(x)$ that satisfy the following
    \begin{align} 
        \label{DTSOSP_constraint1}
        B(x) - \lambda_{\mathcal{X}}(x)s_{\mathcal{X}}(x) &\in \Sigma[x]\\
        \label{DTSOSP_constraint2}
        B(x) - \lambda_{\mathcal{X}_u}(x)s_{\mathcal{X}_u}(x) - 1 &\in \Sigma[x] \\
        \label{DTSOSP_constraint3}
        -B(x) - \lambda_{\mathcal{X}_o}(x)s_{\mathcal{X}_o}(x) + \gamma &\in \Sigma[x]\\
        \nonumber
         -\mathbb{E}[B(F(x, \xi))\ |\ x]  + \frac{B(x)}{\tilde{\alpha}} + \tilde{\beta} -& \\ 
       \label{DTSOSP_constraint4}
        \lambda_{\mathcal{X}_u}(x)s_{\mathcal{X}_u}(x) - \lambda_{\mathcal{X}}(x)s_{\mathcal{X}}(x) &\in \Sigma[x] 
    \end{align}
    where $F(x, \xi) = f(x) + g(x)u(x) + \sigma(x)\xi$ . Then, the probability of failure, depending on the values of $\tilde{\alpha}$ and $\tilde{\beta}$, is defined by (\ref{disbound1}), (\ref{disbound2}) or (\ref{disbound3}).
  \end{theorem}
    We omit the proofs for Theorems \ref{SOSPTheoremCT} and \ref{SOSPTheoremDT}, which follow the general approach for relaxing set constraints to SOS programs using the \emph{Positivstellensatz} condition; see the documentation of \cite{sostools} for details.
    \par 
    For Theorem \ref{SOSPTheoremDT}, the expectation, $\mathbb{E}\big[F(x, \xi) \ | \ x\big]$ in (\ref{DTSOSP_constraint4}) is encoded using the $n$-th moment of a random variable. In the case studies in Section \ref{section: casestudies}, we model the system noise as a random variable with a zero mean normal distribution. The expected value of the $n$-th moment of a normally distributed random variable, $z$, is
    \begin{equation} 
        \mathbb{E}[z^n] = \begin{cases}
        \begin{split}
            \begin{matrix} 0 \ \ &\text{if n is odd} \\ 
            1 \cdot 3 \cdot \cdot \cdot (n - 1)\sigma^n \ &\text{if n is even.}\end{matrix}
        \end{split} \end{cases} 
        \label{eq: expmoment}
    \end{equation}
    Using (\ref{eq: expmoment}) allows for a closed-form expression for $E[B(F(x, \xi)) \ | \ x]$.
\subsection{Verification \& Control Law Synthesis Algorithms}
    \label{subsection: VerControlLawSynth}
    Theorem \ref{SOSPTheoremCT} and \ref{SOSPTheoremDT} are not SOSPs---in fact, they are nonconvex---when all of the relevant parameters are considered variables, i.e., $\alpha$, $\beta$, and $u(x)$. As a result, we present algorithms to numerically compute barrier functions to circumvent the nonconvex problem. Since the algorithms we present are valid for discrete-time and continuous-time systems we use $x$ to represent the continuous-time and discrete-time state instead of $x(t)$ or $x[k]$, respectively. \par
    
    First, we assume that $u(x)$ is fixed, and thus we solve the verification problem via Algorithm \ref{algo: BxalgoCT} which computes a barrier function $B(x)$ satisfying the conditions in Theorem \ref{SOSPTheoremCT}. These conditions are nonconvex in $\alpha$, so we will perform a line search on $\alpha.$  The barrier function is evaluated over the set $\mathcal{X}_0$ and utilized to compute the probability, $P$, using (\ref{bound1}), (\ref{bound2}), or (\ref{bound3}) for continuous-time systems. The polynomial degree $n_B$ of $B(x)$ is a design parameter; however, higher-order polynomials tend to produce tighter bounds. Well refined bounds (i.e. higher-order polynomials) present themselves with the trade-off of longer computational times versus probability of failure refinement. \par
	
	The objective of the SOSP in Algorithm \ref{algo: BxalgoCT} is set to minimize the value $\gamma + \beta$. This objective was chosen to avoid creating bi-linear programs where initialization of the variables can become complex. In other words, minimizing $\gamma + \beta$ is a heuristic which may not be the best but empirically provides reliable performance.
	
    \begin{remark}
        \label{rem:2}
      As in Remark \ref{rem:1}, if $x_0$ is known exactly, $B(x_0)$ can be substituted for $\gamma$  to provide a bound for all initial conditions $x_0\in \mathcal{X}_0$.
    \end{remark}

    The discrete-time procedure follows the general idea of the continuous-time approach and is also presented in Algorithm \ref{algo: BxalgoCT} but the optimization program is instead constrained by (\ref{DTSOSP_constraint1})--(\ref{DTSOSP_constraint4}). Additionally, even though $\tilde{\alpha}$ and $\tilde{\beta}$ appear in Theorem \ref{theorem:SBF_DT}, for clarity, the notation $\alpha$ and $\beta$ is used in Algorithm \ref{algo: BxalgoCT}. Like in continuous-time systems, the discrete-time probability bound of becoming unsafe  is a function of $\tilde{\alpha}$ and $\tilde{\beta}$ and is computed using (\ref{disbound1}), (\ref{disbound2}) or (\ref{disbound3}).
    
    \begin{algorithm} 
        \scriptsize
    	\caption{Compute $B(x)$}
    	\begin{algorithmic}[1] 
    	\Procedure{Compute-$B$}{$l_{\alpha}, u_{\alpha},d, \sigma, u(x), n_B$} 
    	\State \Comment{$\tilde{\alpha} \  \& \  \tilde{\beta}$ used for discrete-time}
    	\State $\alpha \gets Range(l_{\alpha}, u_{\alpha}, d)$ \Comment Assign $\alpha$ values $d$ apart
    	\State $P^* \gets 1$
    	\State $P \gets \emptyset$
    	\For {$\alpha_i \in \alpha$}
    	    \State 
    	    \State \textbf{Continuous-time:}
    		\State $\min$ $ \gamma + \beta$
    		\State $\text{subject to}$  (\ref{CTSOSP_constraint1}) - (\ref{CTSOSP_constraint4})
    		\State 
    		\State Compute $P$, using (\ref{bound1}), (\ref{bound2}) or (\ref{bound3})
    	    \State
    	    \State \textbf{Discrete-time:}
            \State $\min \gamma + \beta$

    		\State $\text{subject to}$ (\ref{DTSOSP_constraint1}) - (\ref{DTSOSP_constraint4})
    		\State 
    		\State Compute $P$, using (\ref{disbound1}), (\ref{disbound2}) or (\ref{disbound3}). 
        	\State 
    		\If{$P < P^*$} 
    			\State $\alpha^* := \alpha_i$
    			\State $\beta^* := \beta$
    			\State $P^* := P$
    		\EndIf
    	\EndFor
    	\State \textbf{return} $\alpha^*, \beta^*, P^*$
    	\EndProcedure
    	\end{algorithmic}
	\label{algo: BxalgoCT}
    \end{algorithm}
    
\subsection{Controller Synthesis Procedure}
    \label{subsect:controlsynthprocedure}
    So far, we have assumed a given feedback control policy $u(x)$. In this section, we will consider the case of solving for $u(x)$ to achieve a desired probability of safety. In general, we synthesize a polynomial feedback control law of the same or lower order of $B(x)$ such that the upper bound on the probability of failure reduces to a designer specified value. First, the polynomial $u(x)$ is written in quadratic form as 
    
    \begin{equation} \label{SMRu}
      u(x) = z^TQz
    \end{equation}
    
    where $z$ is a vector of monomials in $x$ of a specified order and $Q$ is a coefficient matrix of appropriate dimensions.  Because there likely exist many feasible controllers ensuring the desired probability of failure, we introduce a cost criterion to choose among them.
    We approximate the energy of a particular control policy via a proxy measure. In this case, the proxy is the non-negative scalar, $c$, such that the following vector element-wise constraints
    
    \begin{equation} \nonumber
     c\mathds{1} - \vect(Q) \geq 0 
    \end{equation}
    \begin{equation} \nonumber
     \vect(Q) + c\mathds{1}  \geq 0 
    \end{equation}
    
    hold where $\vect(Q)$ is the vector form of matrix $Q$ and $\mathds{1}$ is the vector of ones of appropriate dimension. Constraining the individual values of the polynomial coefficients provides a means of upper-bounding and lower-bounding the control effort applied at each particular state. We choose the cost $\min c$ to minimize the coefficients appearing in the polynomial controller to encourage lower control effort.   This objective and procedure are highlighted in Algorithm \ref{algo: Uxinit}. 
    
    \begin{algorithm}
        \scriptsize
    	\caption{Initialize $u(x)$}
    	\begin{algorithmic}[1]
    		\Procedure{Compute-$u$}{$B(x), \alpha, \beta, n_u$}
        	\State \Comment{$\tilde{\alpha} \  \& \  \tilde{\beta}$ used for discrete-time}
    		\State $u(x) = z^TQz$ \Comment{$u(x)$ is an $n_u$ power polynomial}
    		\State \Comment{$z$ is a vector of state monomials}
    
            \State $\min \ c$
            \State subject to \qquad $c\mathds{1} - \vect(Q) \geq 0 $
            \State \qquad \qquad \qquad $\vect(Q) + c\mathds{1}  \geq 0$ 
            \State \qquad \qquad \qquad \textbf{Continuous-time:} (\ref{CTSOSP_constraint4})
            \State \qquad \qquad \qquad \textbf{Discrete-time:} (\ref{DTSOSP_constraint4})
    		\State \textbf{return} $u(x), c, Q$ 
    		\EndProcedure
    	\end{algorithmic}
    	\label{algo: Uxinit}
    \end{algorithm} \par
    
     Control synthesis is performed using Algorithm \ref{algo:CTcontrolsynthesis} which utilizes the verification approach from Algorithm \ref{algo: BxalgoCT} and interleaves it with the controller search in Algorithm \ref{algo: Uxinit}. Similar to the verification procedure, Algorithm \ref{algo:CTcontrolsynthesis} initially computes a polynomial barrier given a fixed control policy (i.e. $u(x) = 0$). Following this, Algorithm \ref{algo:CTcontrolsynthesis} iteratively synthesizes a feedback control law by adjusting the parameter, $\beta$. Generally speaking, as in our case studies, we are interested in systems where the probability of failure with no control action is above the goal probability and thus control action is required to achieve the desired probability of safety. \par
     
    \begin{algorithm} 
        \scriptsize
    	\caption{Search for control polynomial $u(x)$}
    	\begin{algorithmic}[1]
    		\Procedure{Compute-$u_{goal}$}{$P_{goal},\sigma, \alpha ,n_B, n_u, \epsilon$} 
        	\State \Comment{$\tilde{\alpha} \  \& \  \tilde{\beta}$ used for discrete-time}
    	    \State $i_{count} = 1$ \Comment{Initialize counting variable}
                \While {$|P^* - P_{goal}$ $| > \epsilon $}
    		    \If {$i_{count} = 1$}
    		    	\State $\beta, P \gets$ {COMPUTE- $B$}{$(l_{\alpha}, u_{\alpha}, d, \sigma, u(x),n_B)$}
    		    	\State \Comment{Since $\alpha$ fixed, $l_{\alpha} = u_{\alpha}$}
    		    	\State \Comment{$u(x) = 0$}
    			    \State $i_{count} := i_{count} + 1$
    			\Else 
    						    \State $u(x), c, Q \gets${ COMPUTE-$u$}{($B(x), \alpha, \beta,n_u$)}
    							\State $\beta, P \gets$ {COMPUTE-$B$}{$(l_{\alpha}, u_{\alpha}, d, \sigma, u(x), n_B)$}
    			\EndIf 
    			\State 
    				\If{$P < P_{goal}$ \textbf{\text{and}} $c < c^*$} 
    				\State $\beta^* := \beta$
    				\State $P^* := P$
    				\State $c^* := c$
    				\EndIf
    				\State \Comment{{$c^*$ is initialized as a large number}}
    				\If {$P > P_{goal}$}
    					\State $\beta := a_{dec} \beta$
    				\Else 
    					\State $\beta := a_{inc} \beta$ 
    				\EndIf  
    				\State \Comment{$a_{inc} > 1$ and $a_{dec} < 1$ are scaling factors}
    				\State 
    		\EndWhile
    
    		\State \textbf{return} $u^*(x), c^*, Q$
    		\EndProcedure
    	\end{algorithmic}
        \label{algo:CTcontrolsynthesis}
    \end{algorithm}
    
    The discrete-time procedure for controller synthesis is also demonstrated in Algorithm \ref{algo: Uxinit} and \ref{algo:CTcontrolsynthesis} where $\tilde{\alpha}$ and $\tilde{\beta}$ are utilized instead of $\alpha$ and $\beta$. The objective of the approach we present is to find a control polynomial based on a system's probability of failure. In continuous-time, the condition (\ref{supmartrelax}) is affine-in-control; however, the same is not always true for condition (\ref{expectationevolbound}) of discrete-time systems. In continuous-time systems, the evolution of the expected value is governed by the infinitesimal generator presented in Fact \ref{fact:infinitesimalgen}. In discrete-time, the evolution is governed by the difference between the expected value of the barrier function at $x[k+1]$ and $x[k]$. Since we are considering polynomial barrier functions the search for control polynomials becomes complex due to the $\mathbb{E}[B(F(x, \xi)) \ | \ x]$ term in (\ref{DTSOSP_constraint4}). Because of this, the sum-of-squares program becomes non-linear and is not necessarily convex; however, if the chosen barrier function is linear then the optimization problem remains convex.
    \par 

\section{Case Studies}
    \label{section: casestudies}
    In this section, we first present a simple continuous-time example to illustrate the advantages and limitations of our technique. Second, a nonlinear continuous-time example is presented to demonstrate the versatility of our approach. Lastly, a discrete-time population growth model is considered. For all case studies, we conduct Monte Carlo simulations to establish ground truth probability bounds. We utilize SOSTOOLS \cite{sostools} which converts the SOSP into semidefinite programs. Our choice of solver is the semidefinite program solver SDPT3 \cite{SDTP3_1999, SDTP3}. The noise term in both the continuous-time and discrete-time systems are modeled to be values from a standard normal distribution, $\mathcal{N}(0,1)$. These case studies were conducted on a 2.3 GHz Intel Core i5 computer with 8GB of memory.\footnote{The MATLAB source code for the four case studies is contained at \url{https://github.com/gtfactslab/stochasticbarrierfunctions}}

\subsection{1-D Stochastic System}
\begin{figure} 
    \centering
	\includegraphics[scale=.43]{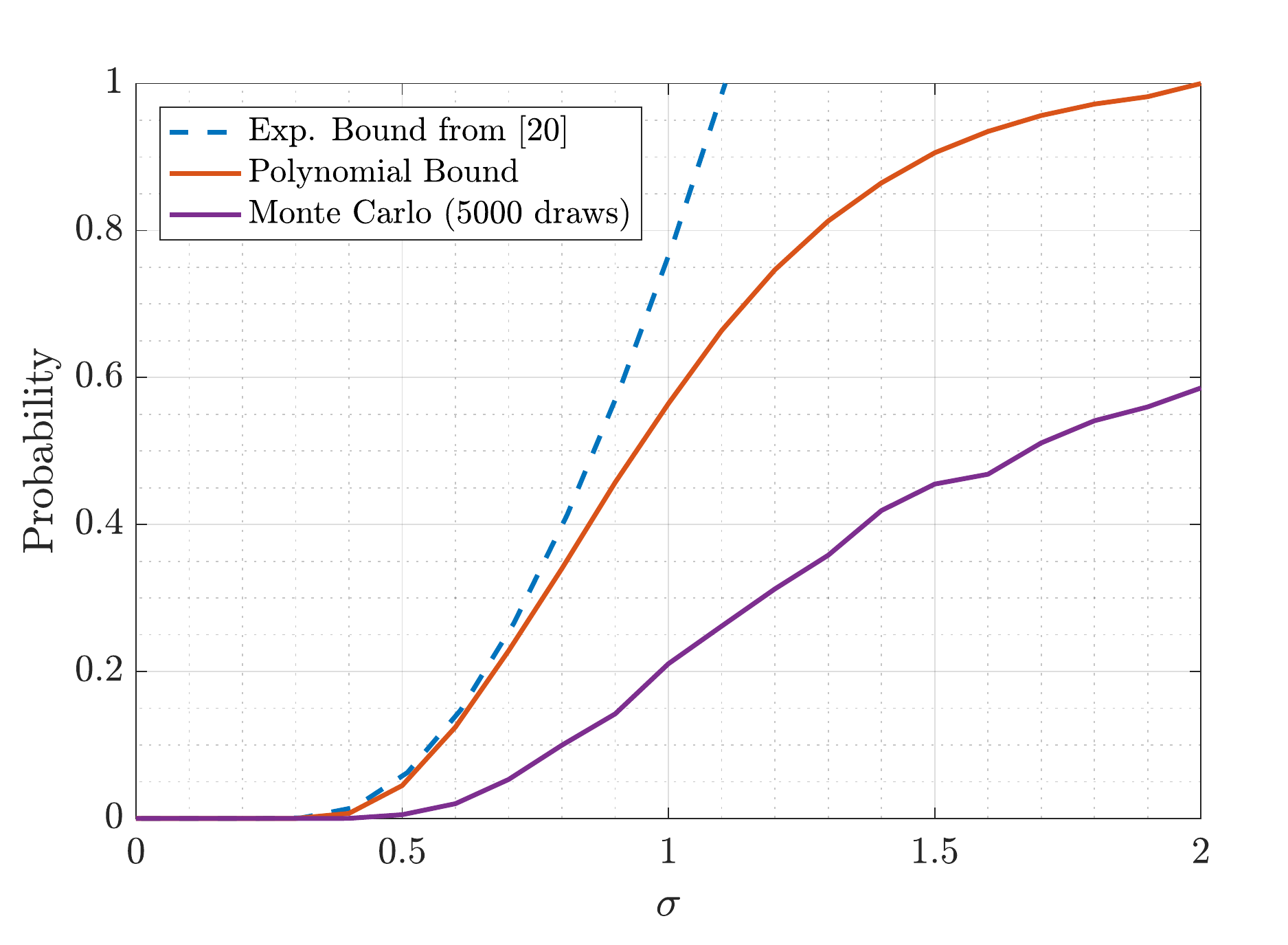}
	\caption{The probability of failure bounds for \eqref{eq: 1Dsystem} are presented here. A 16\textsuperscript{th} degree polynomial barrier function is considered. The Monte Carlo simulation results illustrate the true probability of failure for this system.}
	\label{ex1probbounds}
\end{figure}

     Consider a 1-D stochastic affine-in-control system of the form
    \begin{equation}
        \label{eq: 1Dsystem}
    	dx = \big(-x + u(x)\big)dt + \sigma dw.
    \end{equation}
    This is of the same form as (\ref{controlstochdiff}) where $f(x) = -x$, $g(x)  = 1$, and constant $\sigma(x)\equiv \sigma$. We define the state space as $\mathcal{X} = \{x: -2 \leq x \leq 2 \}$, $\mathcal{X}_u = \{x: x^2 \geq 1\}$, and $\mathcal{X}_0 = \{x: x^2 \leq 0.2^2\}$.
    First, we benchmark the probability of failure without a control input (i.e. $u(x) = 0$) for a finite time horizon of $T = 1 \ \text{s}$. Thus, to do so, the procedure outlined in Algorithm \ref{algo: BxalgoCT} is utilized. We grid search over a defined range of values for the constant $\alpha$. In this particular example, $\alpha \in [0, 5]$ with $d = 0.05$ in Algorithm \ref{algo: BxalgoCT}. We search for a 16\textsuperscript{th} degree $B(x)$. Additionally, the c-martingale bound presented in \cite [Algorithm 3]{SteinhardtJacob2012Frvo} is reproduced. Lastly, the results are benchmarked against the true probability of failure created via a 5000 draw Monte Carlo simulation. The results are presented in Fig. \ref{ex1probbounds}. \par  

In Fig. \ref{ex1probbounds}, the polynomial bound on the probability of failure performs better than the bound from \cite{SteinhardtJacob2012Frvo} generated using the c-martingale condition that is not state-dependent. The difference is particularly notable at higher values of $\sigma$ where the exponential bound from \cite{SteinhardtJacob2012Frvo} becomes trivial, i.e., greater than or equal to one.  \par

Next, the control problem of achieving a particular bound on the probability of failure of this system is addressed. We consider a desired failure probability of $P_{goal} = 0.30$. We restrict our attention to a linear controller of the form $u(x) = -kx$. The search for a low-energy controller which successfully fulfills the design requirement follows a modified binary search version of Algorithm \ref{algo:CTcontrolsynthesis}. This enables a simple search for the $k$ necessary to achieve the desired criterion. \par 

\begin{figure} 
    \centering
	\includegraphics[scale=.45]{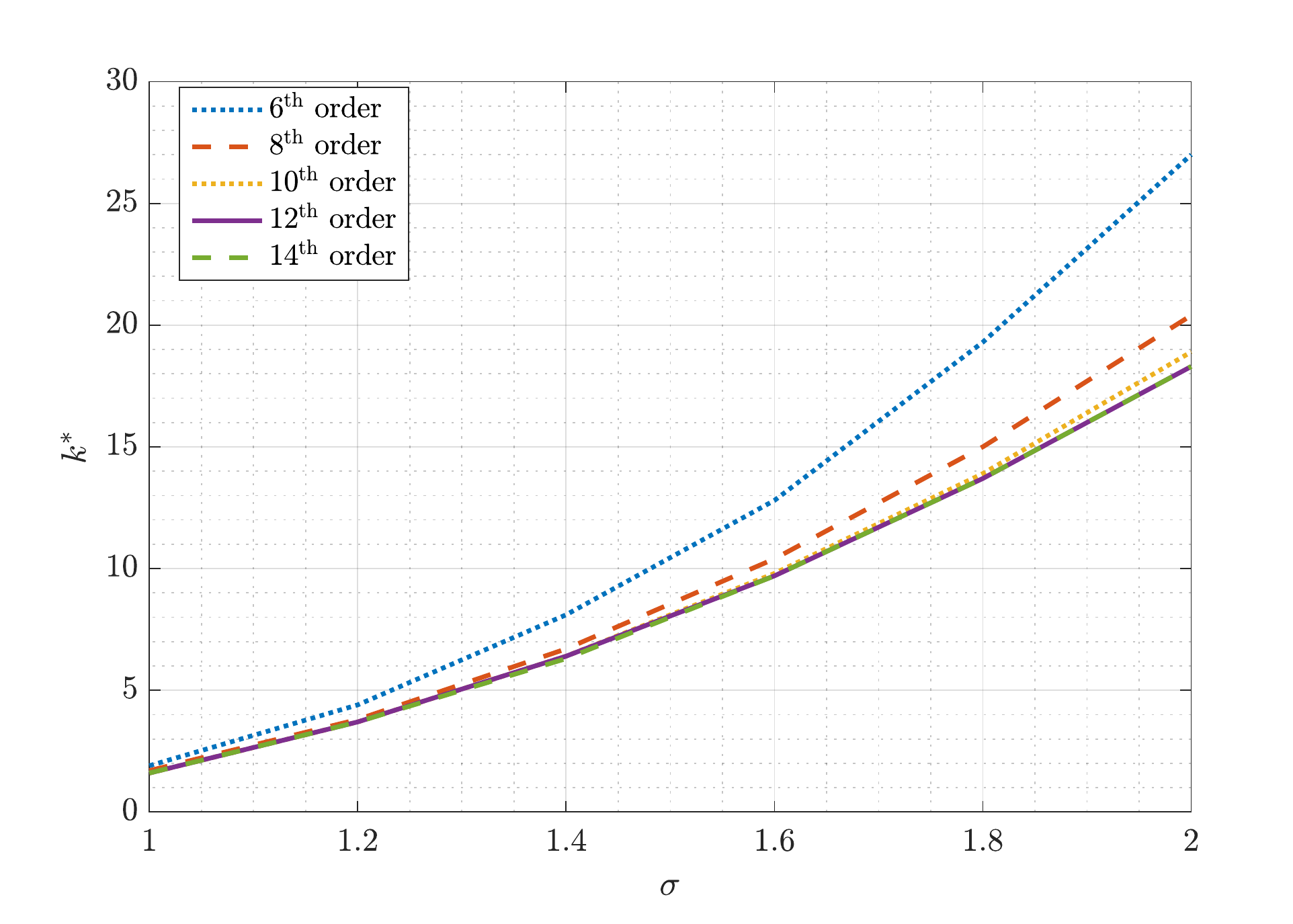}
	\caption{ An illustration of \eqref{eq: 1Dsystem} demonstrating the trade-off between required control gain and the degree of the barrier function, $B(x)$, needed to successfully attain the desired probability of failure threshold. Using higher-order polynomials allows us to guarantee that the desired probability bound is satisfied for a smaller control gain up until some point. Eventually, the order of the polynomial will not improve the bound as is happening from the 12\textsuperscript{th} to 14\textsuperscript{th} order polynomial.}
	\label{ex1controlsearch}
\end{figure}

Fig. \ref{ex1controlsearch} plots $k^*$ achieving the desired failure probability bound for $\sigma \in [1, 2]$. Here, note that the degree of the barrier function for which we search greatly affects the control gain needed to achieve the control objective. In some sense, searching for a higher-order polynomial refines the probability of failure bound requiring lower control effort; however, these high order polynomials require more computation time. Eventually, the degree of the polynomial reaches a saturation point where it does not further decrease the $k^*$ required. 

\subsection{Nonlinear Dynamics}
Consider the stochastic nonlinear dynamics
\begin{align}
    dx_1 &= x_2dt \label{vdp1}  \\
    dx_2 &= \bigg(-x_1 - x_2 - x_1^3+ u(x) \bigg) dt+ \sigma dw.
    \label{vdp2}
\end{align}
This system is studied in \cite{Prajna:2004up} without the input term $u(x)$ and constant $\sigma(x)\equiv \sigma$.
\par
\begin{figure} 
	\centering
	\includegraphics[scale=.45]{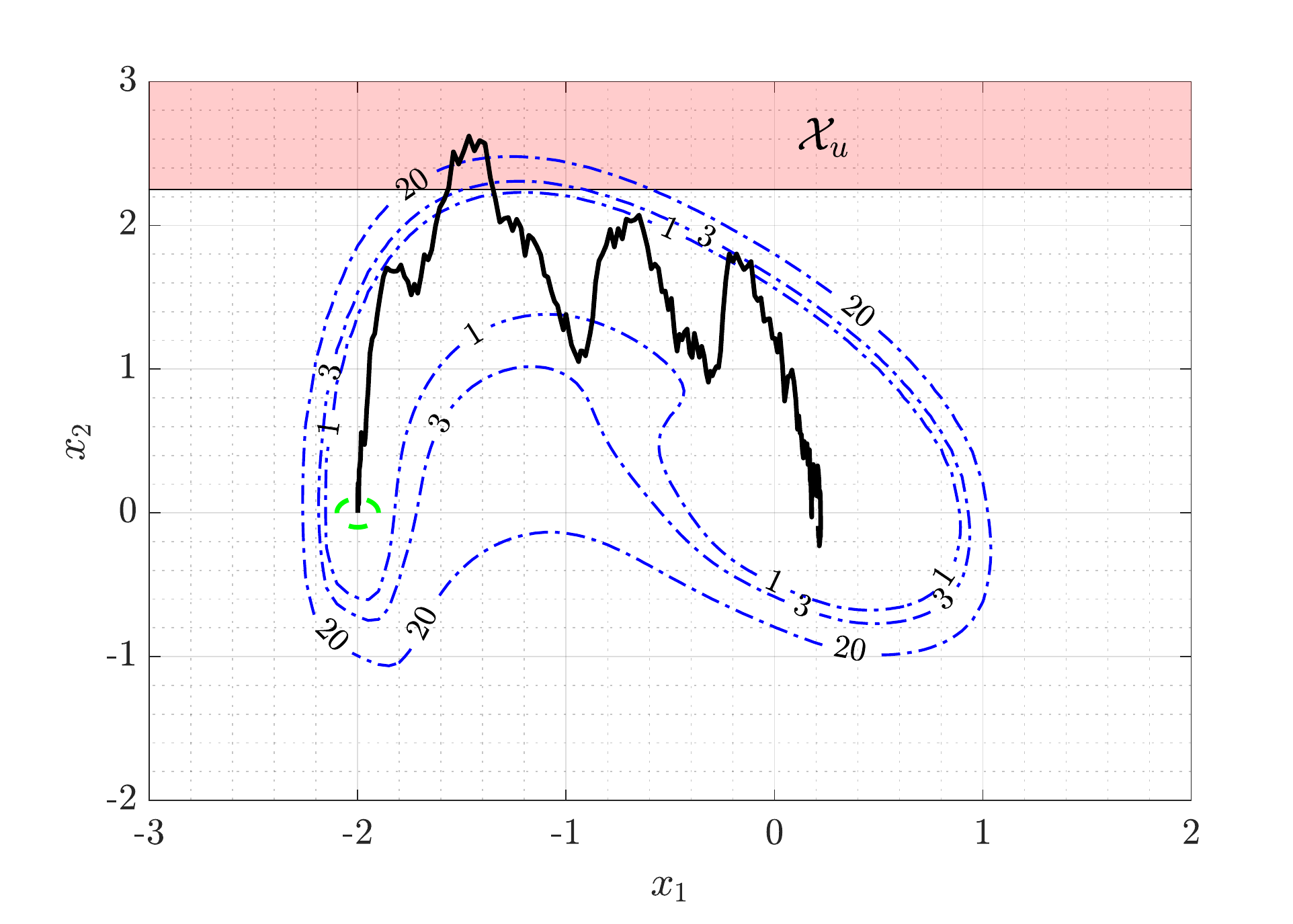}
	\caption{Given the initial conditions $x_0 = [-2, 0]$, the single trajectory dynamics of \eqref{vdp1}--\eqref{vdp2} for a time horizon of $T = 2$ and a $\sigma = 1.0$ are illustrated. The unsafe region is $\mathcal{X}_u = \{x_2 \ | \ x_2 \geq 2.25\}$. Additionally, the level sets of $B(x)$ and their respective values are labeled and given as dashed blue lines.}
	\label{ex2problemillustration}
\end{figure}

 We define the state space as $\mathcal{X} = \{(x_1, x_2) \ | -3 \leq x_1 \leq 2, -2 \leq x_2 \leq 3\}$, $\mathcal{X}_u = \{x_2 \ | \ x_2 \geq 2.25 \}$, and $\mathcal{X}_0 = \{(x_1, x_2)| (x_1 + 2)^2 + x_2^2 \leq 0.1^2 \}$. A sample trajectory of (\ref{vdp1})--(\ref{vdp2}) is illustrated in Fig. \ref{ex2problemillustration}. Additionally, level sets of $B(x)$ are projected onto the state space. In this illustration, $B(x)$ is computed with $u(x) = 0$ solely using Algorithm \ref{algo: BxalgoCT}. \par 
 
\begin{figure} 
	\centering
	\includegraphics[scale=.45]{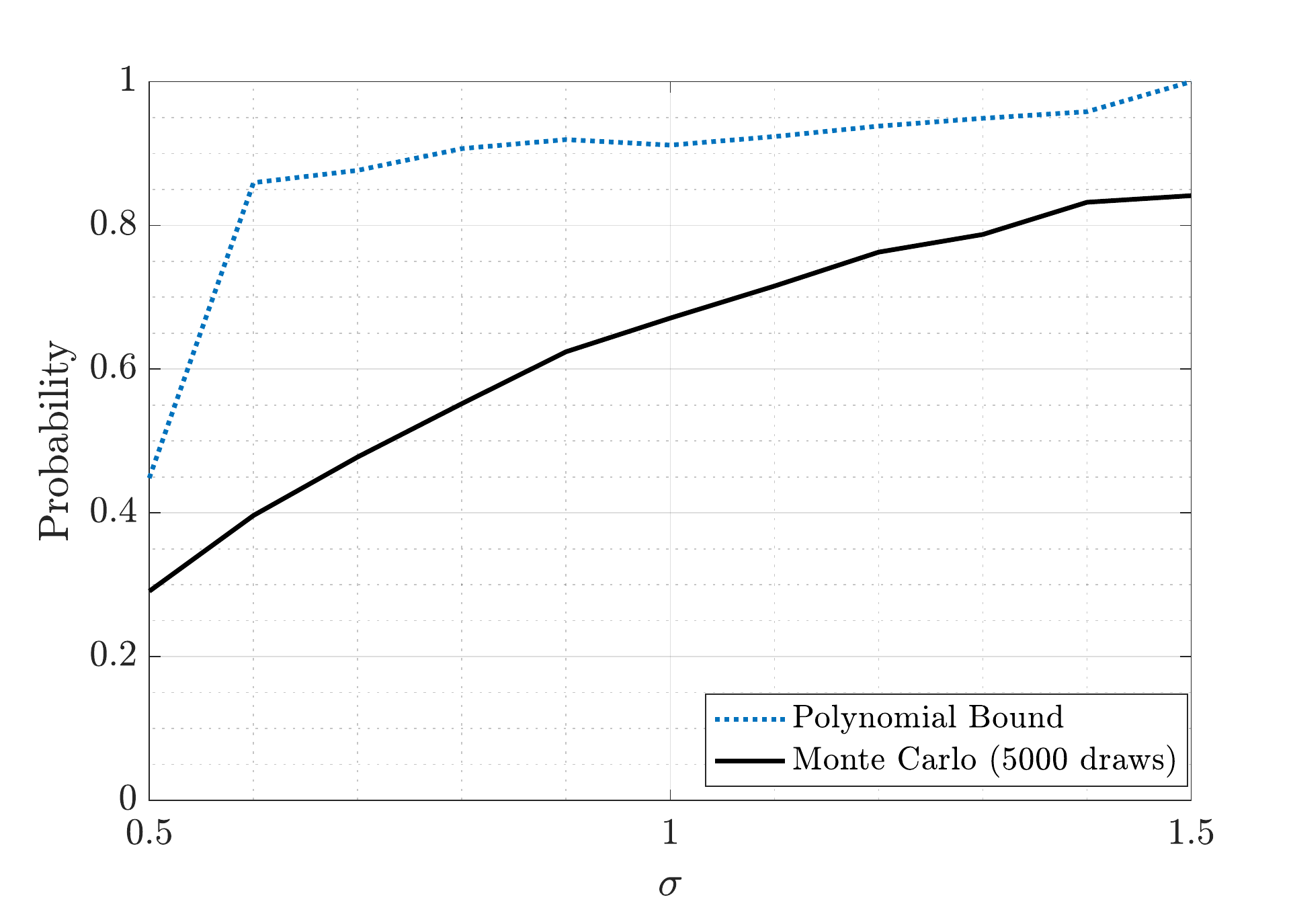}
	\caption{Computing a 14\textsuperscript{th} order polynomial barrier function for the nonlinear dynamics \eqref{vdp1}--\eqref{vdp2}, we are able to bound the probability of failure of the 5000 draw Monte Carlo dynamics for constant $\sigma \in [0.5, 1.5]$.}
	\label{ex2probabilitybound}
\end{figure}

 In this particular trajectory illustration, the evolution of system noise is enough for the system to enter the predefined unsafe set; however, this is not always the case. To illustrate this, we compute a Monte Carlo simulation of the system dynamics shown. Additionally, an upper bound is computed on the probability of becoming unsafe given our initial condition and illustrated in Fig. \ref{ex2probabilitybound}. While a set of initial conditions is encoded into the SOSP, the probability bound is evaluated at the same initial point, $x_0 \in \mathcal{X}_0$, as the Monte Carlo simulation. \par
 
 The feedback control law design specification for this system is to reduce the probability of failure bound to $P_{goal} = 0.10$ for specified $\sigma$ values. For this example a 2\textsuperscript{nd} order polynomial controller of the form of (\ref{SMRu}) is synthesized. The constant, $c$, highlighted in Algorithm \ref{algo: Uxinit} is minimized. Algorithm \ref{algo:CTcontrolsynthesis} produces the results in Table \ref{table: example2controlresult}
 for select values of $\sigma$ and specific $\alpha$ values. The $\alpha$ values in Table \ref{table: example2controlresult} originate from the initial (i.e., $u(x) = 0$) probability bound computation. Here, 10\textsuperscript{th} order $B(x)$ are considered due to the computational limitations of SOSTOOLS.
 
\begin{table}   
    \centering 
    \def\arraystretch{1.35}
         \begin{tabular}    {|| >{\centering\arraybackslash}m{.70in} | >{\centering\arraybackslash}m{.58in}| >{\centering\arraybackslash}m{.58in}| >{\centering\arraybackslash}m{.68in} ||}
            \hline
             $\boldsymbol{\sigma}$ & $\mathbf{P_{u(x) = 0}}$ & ${\boldsymbol{\alpha}}$ &$ \mathbf{\min c}$\\ [0.05ex] 
             \hline\hline
             0.6 & 0.860 & 1.4 & 2.1821\\
             \hline
             0.9 & 0.919 & 1.3 & 0.5251\\
             \hline
             1.0 & 0.912 & 1.3 & 0.6396\\
             \hline
             1.3 & 0.949 & 1.5 & 1.1488 \\
             \hline
    \end{tabular}
    \caption{The results from the search for a control polynomial $u(x)$ which reduces the probability of failure to $P_{goal} = 0.10$ for \eqref{vdp1}--\eqref{vdp2}. The upper-bound on the probability of failure without a given control input is presented here for comparison.}
    \label{table: example2controlresult}
\end{table}

\subsection{Discrete-Time Population Model}
Consider the stochastic version of the discrete-time population growth model from \cite{iannelli2015introduction} 
\begin{align}
        \label{populationdynamics1}
        x_1[k+1] &= m_3 x_2[k] + u(x[k])\\
        \label{populationdynamics2}
        x_2[k+1] &= m_1 x_1[k] + m_2 x_2[k] + \sigma \xi[k]
\end{align}
where $m_1 = 0.5$, $m_2 = 0.95$, and $m_3 = 0.5$. For the discrete time system in (\ref{populationdynamics1})--(\ref{populationdynamics2}), we first perform verification via a polynomial barrier function followed by control synthesis using 1\textsuperscript{st} order barrier functions.

\begin{figure}
    \centering
    \includegraphics[scale= .45]{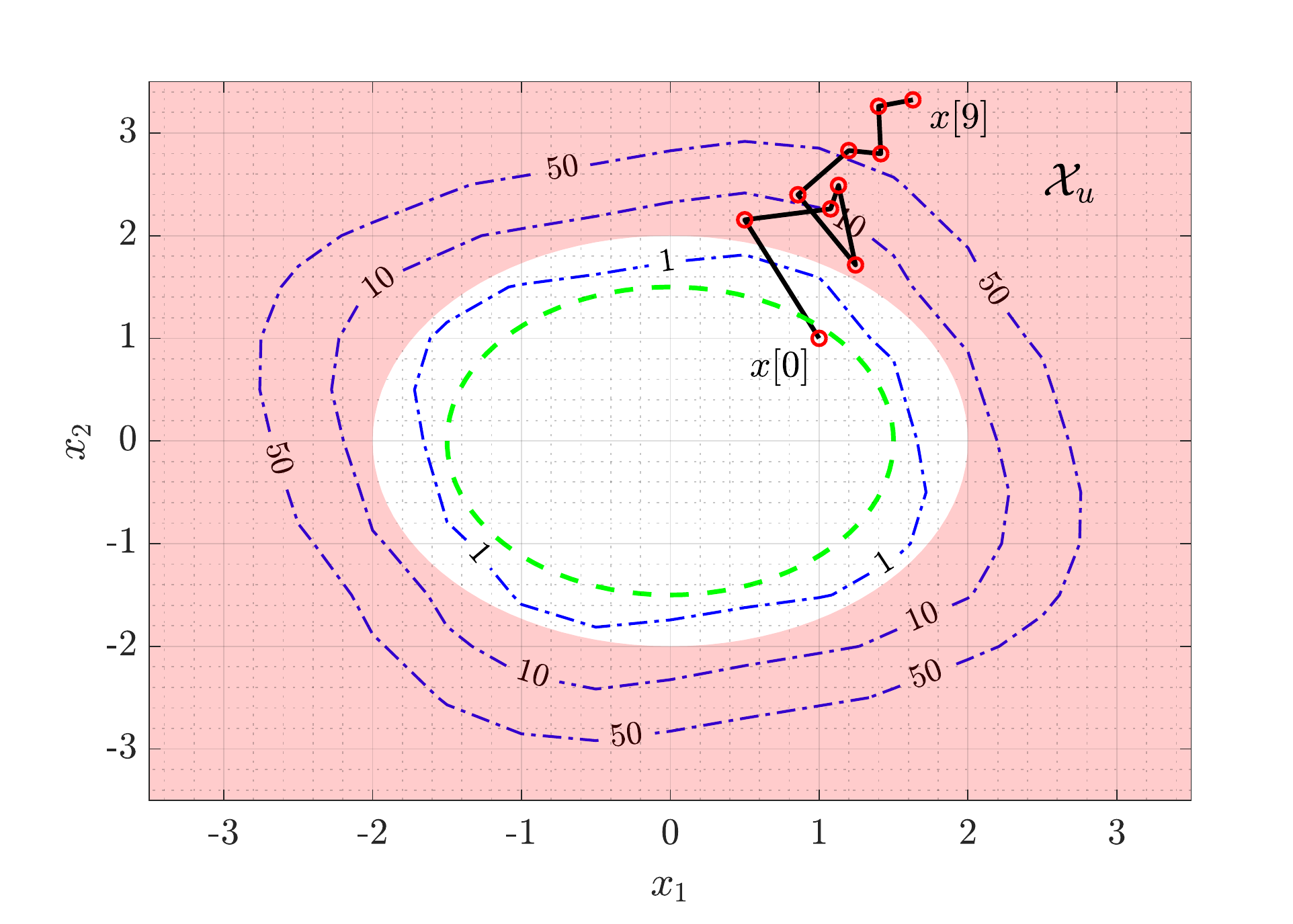}
    \caption{The population dynamics (\ref{populationdynamics1})--(\ref{populationdynamics2}) for $\sigma = 0.5$. The 8\textsuperscript{th} order barrier function, $B(x)$, level sets are superimposed on the state space.}
    \label{fig:cs3_singlerun}
\end{figure}

For verification via polynomial barrier functions, we take $\mathcal{X} = \{x_1, x_2\ | -3 \leq x_1 \leq 3, -3 \leq x_2 \leq 3\}$, $\mathcal{X}_u = \{x_1, x_2\ | \ x_1^2 + x_2^2 \geq 2\}$ and $\mathcal{X}_0 =\{x_1, x_2\ | \ x_1^2 + x_2^2 \leq 1.5 \}$. An illustrative trajectory of the discrete time dynamics (\ref{populationdynamics1})--(\ref{populationdynamics2}) is displayed in Fig. \ref{fig:cs3_singlerun} with the barrier function level sets displayed on the state space. Table \ref{table: DTcontrolresult_popmodelpoly} presents the verification results of Algorithm \ref{algo: BxalgoCT} when $N = 2$ and compares ${P_{u(x[k]) = 0}}$ to the true probability of failure obtained via Monte Carlo simulation for several values of constant $\sigma$. \par
\begin{table}   
    \centering 
    \def\arraystretch{1.35}
     \begin{tabular}    {|| >{\centering\arraybackslash}m{.70in} | >{\centering\arraybackslash}m{.58in}| >{\centering\arraybackslash}m{.58in}| >{\centering\arraybackslash}m{.68in} ||}
        \hline
         $\boldsymbol{\sigma}$ & $\mathbf{P_{u(x[k]) = 0}}$ & \textbf{Monte Carlo} &$\boldsymbol{\gamma}$\\ [0.1ex] 
         \hline\hline
         0.1 & 0.069 & 0.006 & 0.075 \\
         \hline
         0.2 & 0.342 & 0.051 & 0.216 \\
         \hline
         0.3 & 0.574 & 0.118 & 0.261 \\
         \hline
    \end{tabular}
    \caption{Monte Carlo results for the system (\ref{populationdynamics1})--(\ref{populationdynamics2}) and the computed upper bound $P_{u(x[k]) = 0}$ on the probability of failure using an 8\textsuperscript{th} order polynomial. Additionally, the associated $\gamma$ value used to compute the set-wise probability of failure is provided.}\label{table: DTcontrolresult_popmodelpoly}
\end{table}

As highlighted in Section 4.3, in the discrete-time case, evaluating $\mathbb{E}\left[ B(F(x, \xi)) \ | \ B(x)\right]$ results in a nonconvex constraint unless $B(x)$ is affine. Thus, we consider the case when $B(x)$ is affine. We now consider the domain $\mathcal{X} = \{x_1, x_2\ | \ 0\leq x_1\leq 4, 0\leq x_2\leq 4\}$ such that 1\textsuperscript{st} order barrier functions are a viable approach. \par 

\begin{figure}
    \centering
    \includegraphics[scale = .45]{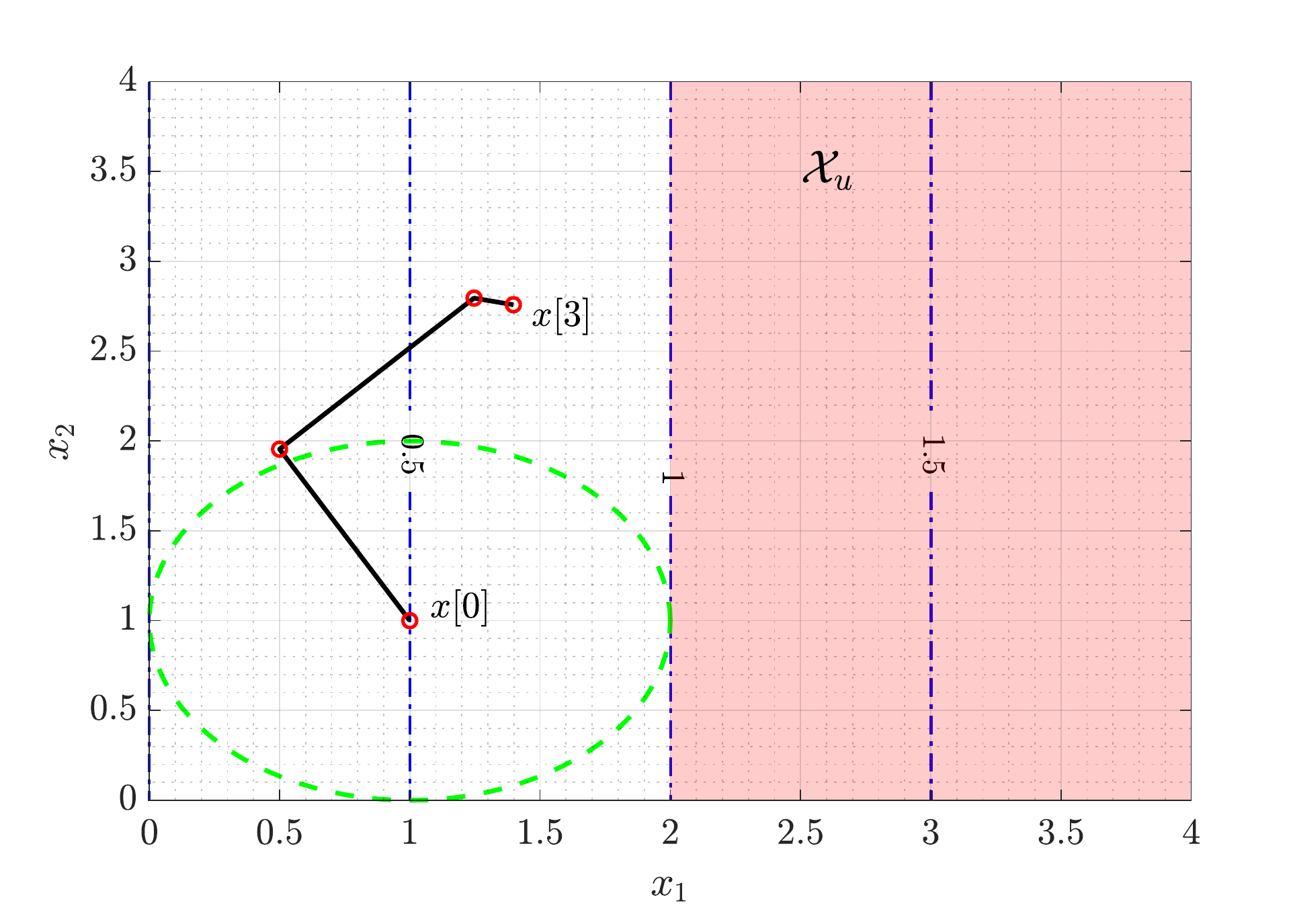}
    \caption{The population dynamics (\ref{populationdynamics1})--(\ref{populationdynamics2}) over a time horizon of N = 3 and $\sigma = 1.5$. The 1\textsuperscript{st} order barrier function, $B(x)$, level sets are super imposed on the state space. Here, we see that the $B(x) \geq 1$ from $x_1 = 2$ to $x_1 = 4$}
    \label{fig:cs3_singlerun_linearbarrier}
\end{figure}

 In the 1\textsuperscript{st} order barrier function case, we take $N = 3$, $P_{goal} = 0.10$ and $\mathcal{X}_u = \{x_1\ | \ 2 \leq x_1 \leq 4\}$. The level sets of a linear barrier function for $\mathcal{X}$ are shown in Fig. \ref{fig:cs3_singlerun_linearbarrier}.
Next, control synthesis for the system is performed using the discrete-time version of Algorithm \ref{algo:CTcontrolsynthesis}. The results of control synthesis are presented in Table \ref{table: DTcontrolresult_popmodel}.
\begin{table}   
    \centering 
    \def\arraystretch{1.35}
     \begin{tabular}    {|| >{\centering\arraybackslash}m{.70in} | >{\centering\arraybackslash}m{.58in}| >{\centering\arraybackslash}m{.58in}| >{\centering\arraybackslash}m{.68in} ||}
        \hline
          $\boldsymbol{\sigma}$ & $\mathbf{P_{u(x[k]) = 0}}$ & $\
         \mathbf{\tilde{\boldsymbol{\alpha}}}$ &$ \mathbf{\min c}$\\ [0.1ex] 
         \hline\hline
         1.0 & 0.499 & 2 & 1.44\\
         \hline
         1.5 & 0.512 & 2.05 & 2.074 \\
         \hline
         2.0 & 0.523 & 2.10 & 2.488\\
         \hline
         2.5 & 0.544 & 2.20 &  2.986\\
         \hline
    \end{tabular}
    \caption{The $c$ value derived from implementing Algorithm \ref{algo:CTcontrolsynthesis} for the system presented in (\ref{populationdynamics1})--(\ref{populationdynamics2}) using a 1\textsuperscript{st} order barrier function for $P_{goal} = 0.10$. The last column gives the value of $c$ which encourages a low-energy control effort for a 2\textsuperscript{nd} order $u(x)$.} \label{table: DTcontrolresult_popmodel}
\end{table}

\section{Conclusion}
\label{section: conclusion}
We consider both continuous-time and discrete-time stochastic control barrier functions whose existence provides a means of quantifying an upper bound on a system's probability of failure. Additionally, we present a novel approach to the problem of finite-time verification by constraining the evolution of the expectation by a non-negative barrier function. This approach includes the supermartingale and c-martingale conditions proposed in prior literature as special cases. Lastly, we synthesize a feedback control strategy $u(x)$ such that a certain probability of failure criterion is met. We illustrate the methods with three case studies which demonstrate our ability to quantify system failure probabilities. For discrete-time systems, we perform verification leveraging polynomial barrier functions; however, controller synthesis in discrete-time systems gives rise to nonconvexities. The discrete-time nonconvexities are mitigated by only considering a region of the state-space such that linear barrier functions are a viable approach using the presented numerical methods. In these case studies, stochastic control barrier functions are synthesized using SOS optimization which enable control synthesis based on the upper-bound on the probability a system will enter an unsafe region of the state space.
\bibliography{references.bib}

\end{document}